# Research Article

# Association between Sitting Time and Urinary Incontinence in the US Population: data from the National Health and Nutrition Examination Survey (NHANES) 2007 to 2018

Guanbo Wang[1] and Xingpeng Di[2,*]

**Affiliation:**

1. Department of Epidemiology, Harvard T.H. Chan School of Public Health, USA

2. Department of Urology, Institute of Urology (Laboratory of Reconstructive Urology), West China Hospital, Sichuan University, Chengdu, China.

**\*Corresponding authors at:**

Prof. Xingpeng Di, Institute of Urology and Department of Urology, West China Hospital, Sichuan University, Chengdu, Sichuan Province, China.

**Word count:** 2569 for plain text; 288 for abstract

**Number of graphs:** 1

**Number of tables:** 3

**Pages:** 15

# Abstract

**Background** Urinary incontinence (UI) is a common health problem that affects the life and health quality of millions of people in the US. We aimed to investigate the association between sitting time and UI.

**Methods** Across-sectional survey of adult participants of National Health and Nutrition Examination Survey 2007-2018 was performed. Weighted multivariable logistic and regression models were conducted to assess the association between sitting time and UI.

**Results** A total of 22916 participants were enrolled. Prolonged sitting time was associated with urgent UI (UUI, Odds ratio [OR] = 1.184, 95% Confidence interval [CI] = 1.076 to 1.302, $P$ = 0.001). Compared with patients with sitting time shorter than 7 hours, moderate activity increased the risk of prolonged sitting time over 7 hours in the fully-adjusted model (OR = 2.537, 95% CI = 1.419 to 4.536, $P$ = 0.002). Sitting time over 7 hours was related to male mixed UI (MUI, OR = 1.581, 95% CI = 1.129 to 2.213, $P$ = 0.010), and female stress UI (SUI, OR = 0.884, 95% CI = 0.795 to 0.983, $P$ = 0.026) in the fully-adjusted model.

**Conclusions** Prolonged sedentary sitting time (> 7 hours) indicated a high risk of UUI in all populations, female SUI and male MUI. Compared with sitting time shorter than 7 hours, the moderate activity could not reverse the risk of prolonged sitting, which warranted further studies for confirmation.

**Keywords:** sedentary sitting; urinary incontinence; physical activity; cross-sectional; National Health and Nutrition Examination Survey

# Background

Urinary incontinence (UI) is a common but underreported disease that affects the life and health quality of approximately 20 million females and 6 million males in the United States (US)[1]. UI has increased the incidence of depression and social isolation[2]. Despite the great progress of therapies targeting UI are available, only a quarter of female receives a diagnosis and clinical care[3]. In general, UI is characterized as involuntary leakage of urine with or without bladder control dysfunction which includes stress urinary incontinence (SUI), urge urinary incontinence (UUI), and mixed urinary incontinence (MUI)[4]. Currently, many risk factors have been reported to contribute to UI. For females, high parity, vaginal delivery history, and menopause were identified as independent risk factors for UI[5, 6]. For males, post-surgery of the prostate (prostatectomy, transurethral resection of prostate) might contribute to the leakage of urine[7]. Importantly, damage to neural control of the bladder and pelvic floor, obesity, age, and others might cause UI in both genders as well[8].

These years, there is increasing focus on sedentary behavior, such as leisure, screen time, and sitting time. Excessive sedentary sitting time has been reported to be associated with several diseases. For instance, a study in Korea demonstrated that prolonged sitting time increased the risk of low back pain[9]. Long duration of sitting was also associated with cardiovascular disease, diabetes mellitus (DM), and depression[10, 11]. In addition, longer sitting time could also elevate the risk of cancer-specific and all-cause mortality[12]. Given the increasing attention and evidence, the World Health Organization's 2020 Global Guidelines on *Physical Activity and Sedentary Behavior*[13] recommends replacing sedentary behavior with moderate physical activities for health keeping[14].

Although a recent study revealed that sedentary behavior duration was related to UUI in older women from the US, there was no study investigating the association between the sitting duration and UI[15]. And scarce evidence was available on sitting time and male UI. Therefore, we performed a cross-sectional study to comprehensively investigate the relationship between sedentary sitting time and UI in both genders from the National Health and Nutrition Examination Survey (NHANES) 2007-2018.

# Methods

## Study population

NHANES (http://www.cdc.gov/nchs/nhanes.htm) is a large survey designed, cross-sectional, and nationally representative for recording the health and nutrition conditions of the US population with two years for each cycle. The interview was performed in homes by well-trained agents. We collected data from 6 cycles (2007 to 2018), including 59842 adults in total. Moreover, 29816 participants were excluded for missing UI and sitting time data. Finally, 22916 participants (10768 female and 12148 male) adults were enrolled for further analysis (Figure 1). Of note, all the protocols were approved by the National Center for Health Statistics ethical review board, and all participants signed the informed consent.

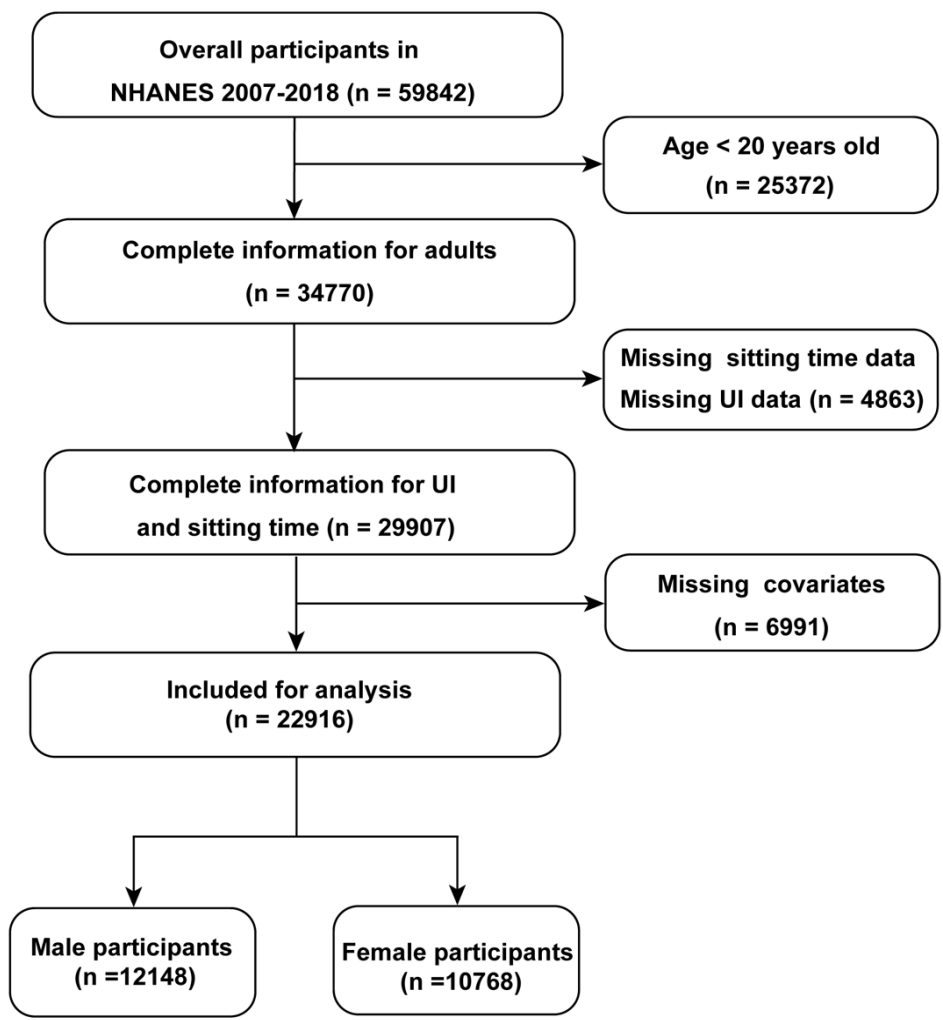

**Figure 1.** Screening of study population. NHANES, National Health and Nutrition Survey; UI, urinary incontinence.

## Definition of Sitting Time

Sitting time is an important component of sedentary activities which is collected by self-report questionnaire. The NHANES defined sitting time as "How much time spent sitting or reclining on a typical day". The sitting time was divided into "< 7 hours" and ">= 7 hours" according to previous studies[9].

## Definition of Urinary Incontinence

The "Kidney Condition-Urology" questionnaire defined SUI by "During the past 12 months, have you leaked or lost control of even a small amount of urine with an activity like coughing, lifting or exercise?". UUI was determined by "During the past 12 months, have you leaked or lost control of even a small amount of urine with an urge or pressure to urinate and you couldn't get to the toilet fast enough like coughing, lifting, or exercise?". MUI was identified as both "yes" answers to SUI and UUI.

## Assessment of Covariates

Multiple covariates were obtained from interviews and examinations, including age, race (Mexican American, non-Hispanic Black, non-Hispanic White, other Hispanic, and other races), education level (Less than 12$^{th}$ grade, High school grade, and College graduate), family income ratio (< 1.3, 1.3-3.5 and > 3.5), marital state, body mass index (BMI, weight/height$^2$, kg/m$^2$), smoking history, alcohol drinking history ( < 1 drink per week, 1-3 drinks per week, and > 4 or more drinks per week), moderate activity, vigorous activity, DM, hypertension, and coronary heart disease.

## Statistical Analysis

The sampling weights, strata, and primary sample units were recommended by the Centers for Disease Control and Prevention (CDC) for better representing the US population. Continuous data were presented as mean ± standard deviation (SD), and categorical variables were presented as numbers with percentages. The differences between groups were compared using the Chi-square test for categorical variables and linear regression for continuous variables.

Furthermore, weighted multivariate logistic regression analyses were adopted to assess the association between sitting time and UI. The crude model was corrected for none. Model 1 was corrected by baseline demographic information, including age, race, education level, family income ratio, and marital state. Besides the confounding factors in model 1, model 2 was further corrected by BMI, smoking history, alcohol drinking history, moderate activity, vigorous activity, DM, hypertension, and coronary heart disease. To clarify the modification effects of covariates, stratified logistic regression analysis was conducted in the UI subgroup. Spline smoothing with a generalized additive model (GAM) was performed to identify the nonlinear relationships.

Statistical analyses were conducted by *R* software version 4.1 (http://www.R-project.org; R Foundation for Statistical Computing, Vienna, Austria) and EmpowerStats (http://www.empowerstats.com, X&Y Solutions, Inc.). A *P* value of < 0.05 (two-sided) indicated statistically significant.

## Results

Among the 22916 participants, 2202 men (weighted proportion 18.13%) and 5860 (weighted proportion 54.42%) were diagnosed as UI (Table 1). The sitting hours were 5.92 ± 3.36 hours in men and 6.04 ± 3.37 hours in women. In women, the most prevalent subtype of UI was SUI, while UUI was in men.

**Table1. Characteristics of participants by categories of gender: NHANES 2007-2018, weighted.**

| Characteristics | Gender | | *P*-value |
| --- | --- | --- | --- |
| | Male | Female | |
| **Number (n)** | 12148 | 10768 | |
| **Age** | 49.76 ± 17.54 | 48.93 ± 17.19 | <0.0001 |
| **Sitting hours (h)** | 5.92 ± 3.36 | 6.04 ± 3.37 | 0.416 |
| **Race (n/%)** | | | <0.0001 |
|   Mexican American | 1798 (14.80%) | 1385 (12.86%) | |
|   Non-Hispanic Black | 2480 (20.41%) | 2313 (21.48%) | |
|   Non-Hispanic White | 5412 (44.55%) | 4979 (46.24%) | |
|   Other Hispanic | 1127 (9.28%) | 1091 (10.13%) | |
|   Other races | 1331 (10.96%) | 1000 (9.29%) | |
| **Education level (n/%)** | | | <0.0001 |
|   Lower than 12th grade | 2921 (24.05%) | 2003 (18.60%) | |
|   High school grade | 2938 (24.19%) | 2340 (21.73%) | |
| College grade | 6289 (51.77%) | 6425 (59.67%) | |
| **Family income ratio (n/%)** | | | 0.002 |
| <1.3 | 3576 (29.44%) | 3317 (30.80%) | |
| >=1.3, <3.5 | 4611 (37.96%) | 4057 (37.68%) | |
| >=3.5 | 3961 (32.61%) | 3394 (31.52%) | |
| **Marital state** | | | <0.0001 |
| married | 6800 (55.98%) | 4870 (45.23%) | |
| Divorced | 1189 (9.79%) | 1556 (14.45%) | |
| Widowed | 472 (3.89%) | 1088 (10.10%) | |
| Separated | 353 (2.91%) | 432 (4.01%) | |
| Living with partner | 1122 (9.24%) | 852 (7.91%) | |
| Never married | 2212 (18.21%) | 1970 (18.29%) | |
| **BMI (kg/m2,n/%)** | | | <0.0001 |
| <=20 | 399 (3.28%) | 633 (5.88%) | |
| >20, <=25 | 2901 (23.88%) | 2647 (24.58%) | |
| >25, <=30 | 4545 (37.41%) | 2966 (27.54%) | |
| >30 | 4303 (35.42%) | 4522 (41.99%) | |
| **Smoking history (n/%)** | | | <0.0001 |
| Non-smoker | 5220 (42.97%) | 6206 (57.63%) | |
| Smoker | 6928 (57.03%) | 4562 (42.37%) | |
| **Alcohol drinking history (drinks/week, n/%)** | | | <0.0001 |
| < 1 | 6925 (57.01%) | 7787 (72.32%) | |
| 1-3 | 3549 (29.21%) | 2267 (21.05%) | |
| >= 4 | 1674 (13.78%) | 714 (6.63%) | |
| **Vigorous activity (n/%)** | | | <0.0001 |
| No | 8873 (73.04%) | 8680 (80.61%) | |
| Yes | 3275 (26.96%) | 2088 (19.39%) | |
| **Moderate activity (n/%)** | | | 0.187 |
| No | 7092 (58.38%) | 6227 (57.83%) | |
| Yes | 5056 (41.62%) | 4541 (42.17%) | |
| **Diabetes mellitus (n/%)** | | | 0.0005 |
| No | 9721 (80.02%) | 8962 (83.23%) | |
| Yes | 2427 (19.98%) | 1806 (16.77%) | |
| **Hypertension (n/%)** | | | 0.004 |
| No | 6778 (55.80%) | 6271 (58.24%) | |
| Yes | 5370 (44.20%) | 4497 (41.76%) | |
| **Coronary heart disease (n/%)** | | | <0.0001 |
| No | 11439 (94.16%) | 10507 (97.58%) | |

|       |           |                    |
|-------|-----------|--------------------|
| Yes   | 709 (5.84%) | 261 (2.42%)      |
| UI    |           |                    | <0.0001
| No    | 9946 (81.87%) | 4908 (45.58%) |
| SUI   | 246 (2.03%) | 2594 (24.09%)   |
| UUI   | 1612 (13.27%) | 1339 (12.43%) |
| MUI   | 344 (2.83%) | 1927 (17.90%)   |

Mean ±SD for continuous variables, P-value was by survey-weighted linear regression. % for categorical variables, *P*-value was by survey-weighted Chi-square test. *P* < 0.05 presents significant difference. h, hour; BMI, body mass index; UI, urinary incontinence; SUI, stress urinary incontinence; UUI. urgency urinary incontinence; MUI, mixed urinary incontinence; NHANES, National Health and Nutrition Examination Survey; h, hour.

The weighted logistic regression analyses were performed to identify the linear relationship between sitting time and UI (Table 2). After being adjusted by demographic characteristics, prolonged sitting time was related to a higher incidence of UUI in males (Odds ratio [OR] = 1.184, 95% Confidence interval [CI] = 1.076 to 1.302, *P* = 0.001). Based on model 1, model 2 showed a similar outcome (OR = 1.114, 95% CI = 1.011 to 1.227, *P* = 0.032). Since the etiologies of UI in both genders were different, we further conducted a gender-stratified analysis. The results demonstrated prolonged sitting time over 7 hours was related to male MUI in the minimally-adjusted model (OR = 1.712, 95% CI = 1.222 to 2.398, *P* = 0.002) and fully-adjusted model (OR = 1.581, 95% CI = 1.129 to 2.213, *P* = 0.010). In addition, prolonged sitting time over 7 hours was related to female SUI in the fully-adjusted model (OR = 0.884, 95% CI = 0.795 to 0.983, *P* = 0.026).

**Table 2. Univariate and multivariate analyses by gender-stratified linear regression model, weighted**.

| UI | Sitting time (h) | All (OR, 95% CI), *P* | Gender Male (OR, 95% CI), *P* | Female (OR, 95% CI), *P* |
|----|------------------|----------------------|-------------------------------|--------------------------|
| UI |                  |                      |                               |                          |
|    | < 7              | Reference            | Reference                     | Reference                |
|    | ≥ 7              |                      |                               |                          |
|    | Crude model      | 1.043 (0.966,1.127), 0.285 | 1.017 (0.997,1.036), 0.099 | 1.001 (0.988,1.014), 0.883 |
|    | Model 1          | 1.055 (0.969,1.149), 0.219 | 1.130 (0.977,1.307), 0.102 | 1.048 (0.943,1.164), 0.386 |
|    | Model 2          | 1.012 (0.929,1.102), 0.789 | 1.075 (0.935,1.237), 0.312 | 0.957 (0.859,1.066), 0.425 |
| **SUI** | | | | |
| | < 7 | Reference | Reference | Reference |
| | ≥ 7 | | | |
| | Crude model | 1.006 (0.928,1.090), 0.883 | 1.045 (1.012,1.079), 0.009 | 0.991 (0.978,1.005), 0.214 |
| | Model 1 | 0.994 (0.909,1.086), 0.888 | 1.317 (1.024,1.695), 0.035 | 0.960 (0.865,1.066), 0.452 |
| | Model 2 | 0.961 (0.878,1.052), 0.383 | 1.243 (0.968,1.598), 0.093 | 0.884 (0.795,0.983), 0.026 |
| **UUI** | | | | |
| | < 7 | Reference | Reference | Reference |
| | ≥ 7 | | | |
| | Crude model | 1.110 (1.018,1.211), 0.020 | 1.015 (0.993,1.037), 0.186 | 1.017 (1.000,1.034), 0.059 |
| | Model 1 | 1.184 (1.076,1.302), 0.001 | 1.158 (0.991,1.352), 0.069 | 1.204 (1.067,1.358), 0.003 |
| | Model 2 | 1.114 (1.011,1.227), 0.032 | 1.101 (0.946,1.282), 0.210 | 1.115 (0.990,1.255), 0.078 |
| **MUI** | | | | |
| | < 7 | Reference | Reference | Reference |
| | ≥ 7 | | | |
| | Crude model | 1.103 (0.978,1.244), 0.114 | 1.064 (1.019,1.111), 0.006 | 1.007 (0.985,1.029), 0.543 |
| | Model 1 | 1.166 (1.017,1.336), 0.030 | 1.712 (1.222,2.398), 0.002 | 1.115 (0.958,1.297), 0.165 |
| | Model 2 | 1.085 (0.952,1.237), 0.227 | 1.581 (1.129,2.213), 0.010 | 1.014 (0.873,1.178), 0.853 |

Crude model: adjusted for none. Model1: adjusted for age, race, education level, family income ratio, and marital state. Model2: adjusted for age, race, education level, family income ratio, marital state, BMI, smoking history, alcohol drinking history, moderate activity, vigorous activity, DM, hypertension, coronary heart disease. $P < 0.05$ presents significant difference. OR, odds ratio; CI, confidence interval; h, hour; BMI, body mass index; DM, diabetes mellitus; UI, urinary incontinence; SUI, stress urinary incontinence; UUI. urgency urinary incontinence; MUI, mixed urinary incontinence.

To confirm the association between sedentary sitting time and UI, a non-linear analysis was performed. Supplementary Figure 1 showed a linear relationship between sitting time and UUI in a fully adjusted model of the whole population of the US. To identify the modifiers for the association between sitting time and MUI in males, a stratified logistic regression analysis was performed. Compared with sitting time within 7 hours, MUI patients tended to be non-Hispanic races with more activity.

Then we further investigate whether the moderate activity could alleviate the association. Compared with patients with sitting time shorter than 7 hours, moderate activity increased the risk of prolonged sitting time over 7 hours in the minimally-adjusted model (OR = 2.708, 95% CI = 1.516 to 4.835, $P$ = 0.001) and fully-adjusted model (OR = 2.537, 95% CI = 1.419 to 4.536, $P$ = 0.002) (Table 3). The non-linear analysis revealed a linear association between MUI and sitting time in men with moderate activities (Supplementary Figure 2). Interestingly, the risk of MUI in prolonged sitting time without moderate activity increased sharply after approximately 12 hours.

**Table 3. Univariate and multivariate analyses by moderate activity-stratified linear regression model, weighted.**

| | Sitting time (h) | |
|---|---|---|
| **Moderate activity** | < 7 (OR, 95% CI) | ≥ 7 (OR, 95% CI) |
| **No** | | |
| Crude model | Reference | 1.547 (1.091,2.193), 0.016 |
| Model 1 | Reference | 1.253 (0.860,1.825), 0.244 |
| Model 2 | Reference | 1.174 (0.827,1.667), 0.373 |
| **Yes** | | |
| Crude model | Reference | 1.961 (1.128,3.412), 0.019 |
| Model 1 | Reference | 2.708 (1.516,4.835), 0.001 |
| Model 2 | Reference | 2.537 (1.419,4.536), 0.002 |

Crude model: adjusted for none. Model1: adjusted for age, race, education level, family income ratio, and marital state. Model2: adjusted for age, race, education level, family income ratio, marital state, BMI, smoking history, alcohol drinking history, vigorous activity, DM, hypertension, coronary heart disease. P < 0.05 presents significant difference. OR, Odds ratio; h, hour; BMI, Body mass index; DM, Diabetes mellitus; h, hour.

## Discussion

Sedentary behavior, was defined as rest or sedentary activities without physical activities consuming energy[16, 17]. Several studies identified an inverse association between sedentary behavior and multiple diseases, including obesity, DM, cardiovascular diseases, and others[18, 19]. As a main component of sedentary behavior, the sitting time has been identified as an important public health issue. Previous studies including 20 nations demonstrated that US adults were sedentary for 7.3 to 7.9 per day[20]. In the current cross-sectional study from the NHANES database, we investigated the association between sedentary sitting time and UI in the adult population of the US. We identified that prolonged sitting time over 7 hours independent risk factor for UUI in all populations, and MUI in men population. The association in men might be modified by race and moderate activity.

Consistent with previous studies, a study investigating the relationship between sedentary behavior and UI in US older women revealed that UUI was significantly related to increased duration of sedentary behavior in women aged 60 years old or older[15]. However, no significant difference was found in the association between sitting time and female UUI. Intriguingly, the difference between sitting time and UUI in both genders was significant. The etiologies of UUI is not clarified to date, and studies reported a link to metabolic syndrome[21]. Given that sedentary behavior is a risk factor for metabolic syndrome, and UUI is prevalent in the obese population, sitting time might be associated with UUI[22, 23]. Interestingly, BMI could not modify the association between sitting time and UI. The outcome was confirmed by a cross-sectional study, reporting that UI was correlated with excess fat mass and poor physical fitness[24].

Furthermore, studies scarcely reported the effect of sitting time on male UI. In our study, we identified the association between sitting time and MUI in males. Sitting more than 7 hours was an independent risk factor for MUI in US men. In addition, moderate activity-stratified analysis demonstrated that prolonged sitting time with moderate activity was associated with a higher risk of MUI. Indeed, the role of physical activity and diseases have not been fully elucidated. Physical activity is beneficial to many diseases, such as DM, obesity, and cardiovascular diseases[25, 26]. The health benefits of "moving more and sitting less" are irrefutable in contemporary society. And studies demonstrated that moderate activity was associated with a lower likelihood of UI in the female population[27]. However, the moderate activity could not reverse the risk of UI in men with prolonged sitting time in our study. Meanwhile, non-linear analysis revealed a rapid increase in the risk of MUI in men without moderate activity. Hence, adequate moderate activity with less sitting time might receive more benefits.

Several biological mechanisms might explain our results. Prolonged sitting time could affect metabolic and sex hormones, inflammation, and immunity process[15]. Furthermore, increasing attention was aroused in sitting-associated abnormal glucose metabolism, which could be attenuated by reducing sitting time[28, 29]. Importantly, studies demonstrated that sitting might diminish the function of moderate activity, which could explain our results in male MUI[30]. Additional studies are needed to evaluate whether physical activities reverse the impairment of prolonged sitting in the UI population.

The strength is that our study is the first to specifically investigate the association between sitting duration and the risk of UI. We identified prolonged sitting time increased the risk of UUI. Furthermore, sedentary sitting for more than 7 hours was also associated with MUI in males. The recommendations of adequate physical activity could attenuate the effects of long-time sedentary activity, encouraging patients to engage more in physical activities. However, limitations could not be avoided. First, the cross-sectional design of NHANES limited further investigation of the effect on UI patients. Second, sedentary sitting time and urine leakage history were obtained by self-report, which might be an origin of bias. Moreover, the physical activity was not clarified into accurate time duration or subtype. More details of the effect of physical activities on UI were not eligible. Finally, the data from NHANES only included the baseline information without behavioral changes in the following time.

## Conclusion

This cross-sectional study showed that prolonged sedentary sitting time (> 7 hours) indicated a high risk of UUI in all populations, female SUI and male MUI. Compared with sitting time shorter than 7 hours, the moderate activity could not reverse prolonged sitting risk, which needs further validation.

## Declarations

## Ethics approval and consent to participate

The studies involving human participants were reviewed and approved by http://www.cdc.gov/nchs/nhanes.htm. Written informed consent to participate in this study was provided by the participant's legal guardian/next of kin.

## Consent for publication

Not applicable.

## Availability of data and materials

All raw data were publicly available in the NHANES database (https://wwwn.cdc.gov/nchs/nhanes/Default.aspx).

## Competing interests

The authors have no conflicts of interest to declare.

## Funding

Not applicable.

## Authors' contributions

BH Liao and XP Di contributed to designing this article, performing the statistical analyses, and drafting the manuscript. XP Di and MH Wang provided the statistical analyses. BH Liao provided critical revision of the manuscript. All authors read and gave final approval of the version to be published.

## Acknowledgements

Thanks to Zhang Jing (Shanghai Tongren Hospital) for his work on the NHANES database. His work on the *nhanesR* package and webpage, make it easier to explore the NHANES database.

## References

1. Irwin GM: **Urinary Incontinence**. *Prim Care* 2019, **46**(2):233-242.
2. Melville JL, Fan M-Y, Rau H, Nygaard IE, Katon WJ: **Major depression and urinary incontinence in women: temporal associations in an epidemiologic sample**. *Am J Obstet Gynecol* 2009, **201**(5):490.e491-490.e497.
3. Lukacz ES, Santiago-Lastra Y, Albo ME, Brubaker L: **Urinary Incontinence in Women: A Review**. *JAMA* 2017, **318**(16):1592-1604.
4. Abrams P, Cardozo L, Fall M, Griffiths D, Rosier P, Ulmsten U, van Kerrebroeck P, Victor A, Wein A: **The standardisation of terminology of lower urinary tract function: report from the Standardisation Sub-committee of the International Continence Society**. *Neurourol Urodyn* 2002, **21**(2):167-178.
5. Wilson PD, Herbison RM, Herbison GP: **Obstetric practice and the prevalence of urinary incontinence three months after delivery**. *Br J Obstet Gynaecol* 1996, **103**(2):154-161.
6. Thom DH, van den Eeden SK, Brown JS: **Evaluation of parturition and other reproductive variables as risk factors for urinary incontinence in later life**. *Obstet Gynecol* 1997, **90**(6):983-989.
7. Pearlman A, Kreder K: **Evaluation and treatment of urinary incontinence in the aging male**. *Postgrad Med* 2020, **132**(sup4).

8.  Wyndaele JJ, Kovindha A, Madersbacher H, Radziszewski P, Ruffion A, Schurch B, Castro D, Igawa Y, Sakakibara R, Wein A: **Neurologic urinary incontinence**. *Neurourol Urodyn* 2010, **29**(1):159-164.
9.  Park S-M, Kim H-J, Jeong H, Kim H, Chang B-S, Lee C-K, Yeom JS: **Longer sitting time and low physical activity are closely associated with chronic low back pain in population over 50 years of age: a cross-sectional study using the sixth Korea National Health and Nutrition Examination Survey**. *Spine J* 2018, **18**(11):2051-2058.
10. Bailey DP, Hewson DJ, Champion RB, Sayegh SM: **Sitting Time and Risk of Cardiovascular Disease and Diabetes: A Systematic Review and Meta-Analysis**. *Am J Prev Med* 2019, **57**(3):408-416.
11. Pavey TG, Brown WJ: **Sitting time and depression in young women over 12-years: The effect of physical activity**. *J Sci Med Sport* 2019, **22**(10):1125-1131.
12. O'Rourke K: **Higher sitting time increases the risk of all-cause, cancer-specific, and noncancer mortality**. *Cancer* 2022, **128**(9):1722.
13. Bull FC, Al-Ansari SS, Biddle S, Borodulin K, Buman MP, Cardon G, Carty C, Chaput J-P, Chastin S, Chou R *et al*: **World Health Organization 2020 guidelines on physical activity and sedentary behaviour**. *Br J Sports Med* 2020, **54**(24):1451-1462.
14. Cao C, Friedenreich CM, Yang L: **Association of Daily Sitting Time and Leisure-Time Physical Activity With Survival Among US Cancer Survivors**. *JAMA Oncol* 2022, **8**(3):395-403.
15. Jerez-Roig J, Booth J, Skelton DA, Giné-Garriga M, Chastin SFM, Hagen S: **Is urinary incontinence associated with sedentary behaviour in older women? Analysis of data from the National Health and Nutrition Examination Survey**. *PLoS One* 2020, **15**(2):e0227195.
16. Biswas A, Oh PI, Faulkner GE, Bajaj RR, Silver MA, Mitchell MS, Alter DA: **Sedentary time and its association with risk for disease incidence, mortality, and hospitalization in adults: a systematic review and meta-analysis**. *Ann Intern Med* 2015, **162**(2):123-132.
17. Owen N, Healy GN, Matthews CE, Dunstan DW: **Too much sitting: the population health science of sedentary behavior**. *Exerc Sport Sci Rev* 2010, **38**(3):105-113.
18. Howard RA, Freedman DM, Park Y, Hollenbeck A, Schatzkin A, Leitzmann MF: **Physical activity, sedentary behavior, and the risk of colon and rectal cancer in the NIH-AARP Diet and Health Study**. *Cancer Causes Control* 2008, **19**(9):939-953.
19. Nam JY, Kim J, Cho KH, Choi Y, Choi J, Shin J, Park E-C: **Associations of sitting time and occupation with metabolic syndrome in South Korean adults: a cross-sectional study**. *BMC Public Health* 2016, **16**:943.
20. Matthews CE, Chen KY, Freedson PS, Buchowski MS, Beech BM, Pate RR, Troiano RP: **Amount of time spent in sedentary behaviors in the United States, 2003-2004**. *Am J Epidemiol* 2008, **167**(7):875-881.
21. Bunn F, Kirby M, Pinkney E, Cardozo L, Chapple C, Chester K, Cruz F, Haab F, Kelleher C, Milsom I *et al*: **Is there a link between overactive bladder and the metabolic syndrome in women? A systematic review of observational studies**. *Int J Clin Pract* 2015, **69**(2):199-217.

22. Edwardson CL, Gorely T, Davies MJ, Gray LJ, Khunti K, Wilmot EG, Yates T, Biddle SJH: **Association of sedentary behaviour with metabolic syndrome: a meta-analysis**. *PLoS One* 2012, **7**(4):e34916.
23. Lawrence JM, Lukacz ES, Liu I-LA, Nager CW, Luber KM: **Pelvic floor disorders, diabetes, and obesity in women: findings from the Kaiser Permanente Continence Associated Risk Epidemiology Study**. *Diabetes Care* 2007, **30**(10):2536-2541.
24. Moreno-Vecino B, Arija-Blázquez A, Pedrero-Chamizo R, Alcázar J, Gómez-Cabello A, Pérez-López FR, González-Gross M, Casajús JA, Ara I: **Associations between obesity, physical fitness, and urinary incontinence in non-institutionalized postmenopausal women: The elderly EXERNET multi-center study**. *Maturitas* 2015, **82**(2):208-214.
25. Smith AD, Crippa A, Woodcock J, Brage S: **Physical activity and incident type 2 diabetes mellitus: a systematic review and dose-response meta-analysis of prospective cohort studies**. *Diabetologia* 2016, **59**(12):2527-2545.
26. González-Gross M, Meléndez A: **Sedentarism, active lifestyle and sport: Impact on health and obesity prevention**. *Nutr Hosp* 2013, **28 Suppl 5**:89-98.
27. Kim MM, Ladi-Seyedian S-S, Ginsberg DA, Kreydin EI: **The Association of Physical Activity and Urinary Incontinence in US Women: Results from a Multi-Year National Survey**. *Urology* 2022, **159**:72-77.
28. Dogra S, Wolf M, Jeffrey MP, Foley RCA, Logan-Sprenger H, Jones-Taggart H, Green-Johnson JM: **Disrupting prolonged sitting reduces IL-8 and lower leg swell in active young adults**. *BMC Sports Sci Med Rehabil* 2019, **11**:23.
29. Dunstan DW, Dogra S, Carter SE, Owen N: **Sit less and move more for cardiovascular health: emerging insights and opportunities**. *Nat Rev Cardiol* 2021, **18**(9):637-648.
30. Akins JD, Crawford CK, Burton HM, Wolfe AS, Vardarli E, Coyle EF: **Inactivity induces resistance to the metabolic benefits following acute exercise**. *J Appl Physiol (1985)* 2019, **126**(4):1088-1094.

**Figure Legends**

**Figure 1.** Screening of study population. NHANES, National Health and Nutrition Survey; UI, urinary incontinence.

# Supplementary Material

**Supplementary Figure 1.** Smooth curve fitting between sitting time and UUI. UUI, urgent urinary incontinence.

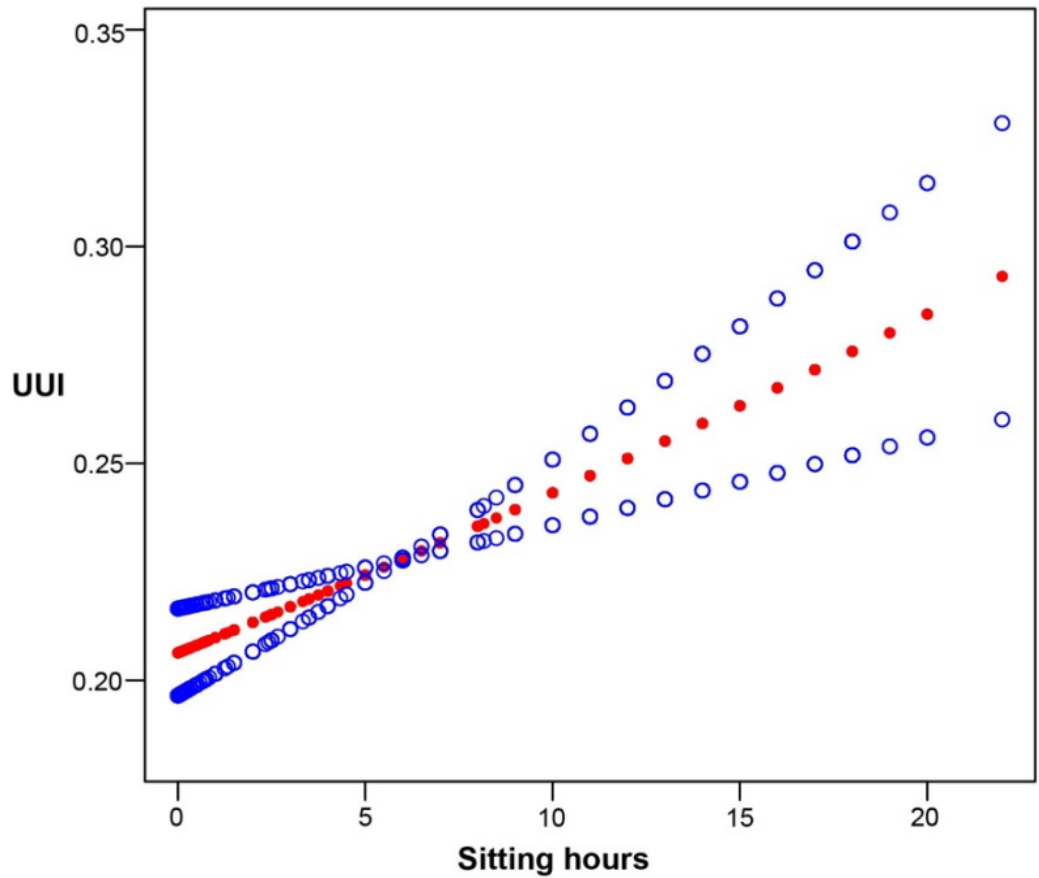

**Supplementary Figure 2.** Smooth curve fitting between MUI and sitting time in men with moderate activities. MUI, mixed urinary incontinence.

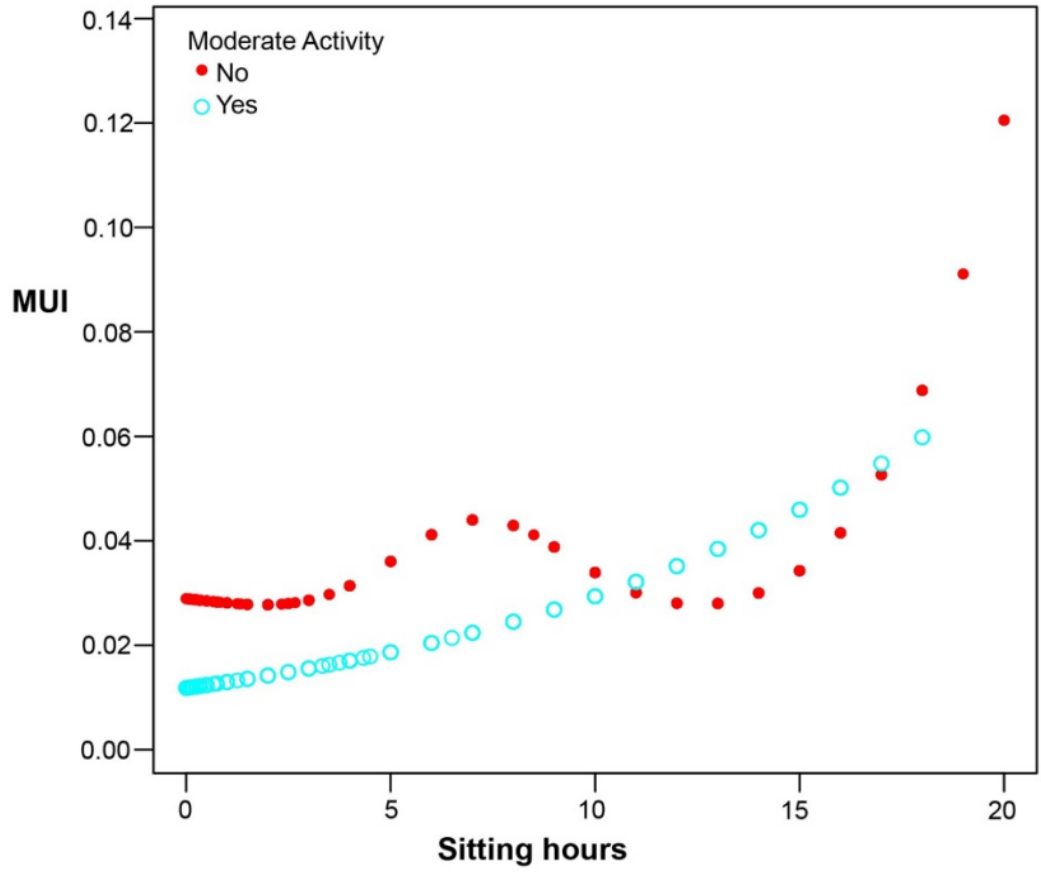